\documentclass[prl,superscriptaddress,showpacs,longbibliography,reprint]{revtex4-2}

\usepackage[colorlinks=true,linkcolor=blue,urlcolor=blue,citecolor=blue,pdfusetitle]{hyperref}
\usepackage{color}
\usepackage[utf8]{inputenc}
\usepackage[usenames,dvipsnames]{xcolor}
\usepackage{amsmath}
\usepackage{amssymb}
\usepackage{graphicx}
\graphicspath{{graphics/}}
\usepackage{epsfig}
\usepackage{dcolumn}
\usepackage{bm}
\usepackage{mathrsfs}
\usepackage{multirow}
\usepackage[all]{xy}
\usepackage{pbox}
\usepackage{lipsum}
\usepackage{verbatim}
\usepackage{braket}
\usepackage{dsfont}
\usepackage{array}
\usepackage{makecell}
\usepackage{tabularx}
\usepackage{isotope}
\usepackage{xr}
\usepackage[normalem]{ulem}

\usepackage[colorlinks=true,linkcolor=blue,urlcolor=blue,citecolor=blue,pdfusetitle]{hyperref}

\begin{document}

\title{Exploiting nonequilibrium phase transitions and strong symmetries for continuous measurement of collective observables
}
\author{Albert Cabot}
\affiliation{Institut f\"ur Theoretische Physik, Eberhard Karls Universit\"at T\"ubingen, Auf der
Morgenstelle 14, 72076 T\"ubingen, Germany.}
\author{Federico Carollo}
\affiliation{Institut f\"ur Theoretische Physik, Eberhard Karls Universit\"at T\"ubingen, Auf der
Morgenstelle 14, 72076 T\"ubingen, Germany.}
\author{Igor Lesanovsky}
\affiliation{Institut f\"ur Theoretische Physik, Eberhard Karls Universit\"at T\"ubingen, Auf der
Morgenstelle 14, 72076 T\"ubingen, Germany.}
\affiliation{School of Physics and Astronomy, University of Nottingham, Nottingham, NG7 2RD, UK.}
\affiliation{Centre for the Mathematics and Theoretical Physics of Quantum Non-Equilibrium Systems,
University of Nottingham, Nottingham, NG7 2RD, UK}

\begin{abstract}
Dissipative many-body quantum
dynamics can feature strong symmetries which give rise to conserved quantities. We discuss here how a strong symmetry in conjunction with a nonequilibrium phase transition allows to devise a protocol for measuring collective many-body observables. To demonstrate this idea we consider a collective spin system whose constituents are governed by a dissipative dynamics that conserves the total angular momentum. We show that by continuously monitoring the system output the value of the total angular momentum can be inferred directly from the time-integrated emission signal, without the need of repeated projective measurements or reinitializations of the spins. This may offer a route towards the measurement of collective properties in qubit ensembles, with applications in quantum tomography, quantum computation and quantum metrology.
\end{abstract}
\maketitle

{\it Introduction. --} Nonequilibrium phases emerge in many-body quantum systems due to the combined action of driving, dissipation and interactions \cite{Diehl2010,Jin2013,Lee2013,Marcuzzi2016,Rota2019}. Dissipative or nonequilibrium phase transitions manifest as nonanalytic changes in the stationary state of the system  \cite{Kessler2012,Minganti2018}  and are accompanied by a rich phenomenology such as long relaxation times and intermittency \cite{Lee2012,Ates2012,Lesanovsky2013,Rose2016,Casteels2017,Fitzpatrick2017,Fink2018,Minganti2023}, squeezing and quantum correlations \cite{Rota2017,Hannukainen2018,Hwang2018,Buonaiuto2021}, or spectral singularities \cite{Minganti2018,Cabot2024b,Dutta2024}. The sharp change occurring near a transition point also constitutes  a resource for sensing and metrology, generally allowing for enhanced sensitivity  \cite{Macieszczak2016,Fernandez2017,Heugel2019,Garbe2020,Ilias2022,Ding2022,Gandia2023,Pavlov2023,Montenegro2023,Ilias2024,Hotter2024}. In quantum optical settings, such as atomic ensembles interfaced with optical cavities or waveguides, the nature of the stationary state manifests in properties of the emitted light \cite{Klinder2015,Muniz2020,Ferioli2023}. This allows to study phases and phase transitions through continuous monitoring of their output, as in photocounting or homodyne detection experiments  \cite{Link2019,Ilias2022,Cabot2023},  and to further exploit this to devise parameter estimation protocols \cite{Ilias2022,Gambetta2001,Gammelmark2013,Kiilerich2014,Kiilerich2016,Albarelli2017,Albarelli2018,Shankar2019,Albarelli2020,Rossi2020,Fallani2022,Nurdin2022,Godley2023,Yang2023,Cabot2024}.

\begin{figure}[t!]
 \centering
 \includegraphics[width=0.9\columnwidth]{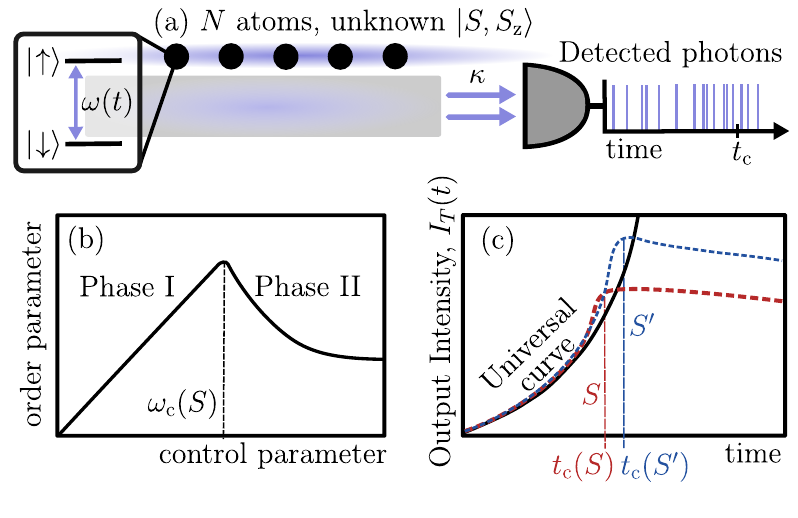}
 \caption{{\bf Protocol.} (a) We consider a spin system realized by $N$ two-level atoms, driven resonantly with Rabi frequency $\omega(t)$ and with collective emission rate $\kappa$, in which the total angular momentum $S$ is a conserved quantity. Collective emission can be realised with, e.g., atoms coupled to a waveguide \cite{Buonaiuto2021} or a cavity mode \cite{Kessler2021,Kongkhambut2022}. The atoms are initially prepared in an unknown total angular momentum state $|S,S_\mathrm{z}\rangle$ and their emission is continuously monitored.  (b) The system displays two distinct dynamical regimes separated by a sharp crossover (phase transition in the thermodynamic limit). The Rabi frequency at which the sharp change occurs depends on $S$, i.e. $\omega_\mathrm{c}(S)$. (c) Adiabatically ramping up the Rabi frequency and monitoring the number of detected photons, one observes a change in the emission statistics at time $t_\mathrm{c}(S)$ due to the $S$-dependent crossover around $\omega_\mathrm{c}(S)$. For $\omega<\omega_\mathrm{c}(S)$ the statistics is universal, and the number of detected photons follows the same curve independently of $S$.} 
 \label{fig_cartoon}
\end{figure}

Open quantum systems can feature  so-called strong symmetries \cite{Buca2012,Albert2014}, which may correspond to physical quantities, e.g. total angular momentum, that are preserved during the time evolution. Strong symmetries have been identified in collective spin systems \cite{Munoz2019,Iemini2024,Solanki2024}, atomic lattices \cite{Tindall2019,Riera2020,Dutta2020,Buca2022,Alaeian2022,Halati2022,Tindall2023} and nonlinear resonators \cite{Lieu2020,Gravina2023,Minganti2023,Labay2024}, for which applications in quantum information have been proposed \cite{Lieu2020,Gravina2023,Labay2024}. A system with a strong symmetry possesses multiple stationary states, one for each symmetry sector \cite{Buca2012,Albert2014}.  Consequently, nonequilibrium transitions can occur independently within each sector and the corresponding critical points  may be located at different parameter values \cite{Halati2022,Iemini2024,Solanki2024}. At the level of single realizations of a continuous monitoring, {dissipative freezing} can occur, in which an initial superposition state living on  different symmetry sectors collapses into a single sector \cite{Munoz2019,Manzano2021,Halati2022,Tindall2023}. The strong symmetry, which is present in the average dynamics,  can thus be violated in {\it single} quantum trajectories. 

In this work, we demonstrate the potential of symmetry-dependent phase transitions in  sensing applications. In particular, we discuss  how the continuous monitoring of the system output allows to infer the specific value assumed by the strong symmetry of the system. To illustrate this approach, we show how to estimate the total angular momentum of an ensemble of two-level atoms, see Fig.~\ref{fig_cartoon}, undergoing a dissipative dynamics  governed by a generator featuring a strong symmetry. We estimate the achievable sensitivity and also consider experimental aspects, such as finite photo-detection efficiency and local (single-atom) decay.

{\it Collective spin system. --} We consider the dissipative dynamics of a system described by a Markovian master equation ($\hbar=1$):
\begin{equation}\label{ME}
\partial_t \hat{\rho}=-i[\hat{H},\hat{\rho}]+\sum_\alpha (\hat{L}_\alpha \hat{\rho}\hat{L}_\alpha^\dagger-\frac{1}{2}\{\hat{L}^\dagger_\alpha \hat{L}_\alpha, \hat{\rho}\})\, .
\end{equation}
Here, $\hat{\rho}$ is the state of the system, $\hat{H}$ its Hamiltonian and $\hat{L}_\alpha$ the jump operators. An Hermitian operator $\hat{A}$ that commutes both with the Hamiltonian and all jump operators $[\hat{H},\hat{A}]=[\hat{L}_\alpha,\hat{A}]=0$, $\forall \alpha$, is known as a strong symmetry  of the system \cite{Buca2012,Albert2014}. In the presence of such an operator, Eq.~(\ref{ME}) features a block diagonal structure and possesses a stationary state for each eigenvalue of $\hat{A}$. The system we consider consists of $N$ spin-$1/2$ particles described by:
\begin{equation}\label{ME_model1}
\hat{H}=\omega\hat{S}_\mathrm{x},\quad \hat{L}=\sqrt{\kappa} \hat{S}_-.
\end{equation}
The collective angular momentum operators are defined as $\hat{S}_\mathrm{\alpha}=\frac{1}{2}\sum_{j=1}^N \hat{\sigma}_\mathrm{\alpha}^{(j)}$ ($\alpha=\mathrm{x,y,z}$), with $\hat{\sigma}_\mathrm{\alpha}^{(j)}$ being the Pauli matrices and $\hat{S}_\pm=\hat{S}_\mathrm{x}\pm i\hat{S}_\mathrm{y}$. This model encodes collective spin decay with rate $\kappa$ and a (resonant) driving with Rabi frequency $\omega$. The total angular momentum  $\hat{S}^2$ is a strong symmetry, making  the use of total angular momentum states convenient. These satisfy $\hat{S}^2|S,S_\mathrm{z},i\rangle=S(S+1)|S,S_\mathrm{z},i\rangle$, and $\hat{S}_\mathrm{z}|S,S_\mathrm{z},i\rangle=S_\mathrm{z}|S,S_\mathrm{z},i\rangle$ with $S=0,1,\dots,N/2$ (for even $N$) \cite{Chase2008,Bargiola2010}. The label $i$ distinguishes the degenerate irreducible representations, or sectors, for each $S$ at a given $N$. We consider initial states belonging to one of these sectors. Hence, the Hilbert space can be simply labeled by $|S,S_\mathrm{z}\rangle$ \cite{Chase2008,Bargiola2010}.

The considered dynamics [cf.~Eq.~(\ref{ME_model1})] within a single sector  features a crossover between two dynamical regimes separated at $\omega_\mathrm{c}(S)=\kappa S$ \cite{Hannukainen2018,Carmichael1980}, see Fig.~\ref{fig_cartoon}(b). For $\omega<\omega_\mathrm{c}(S)$, it displays an overdamped decay toward an almost pure stationary state. For $\omega>\omega_\mathrm{c}(S)$, it displays long-lived oscillations and eventually approaches a highly mixed state. The crossover gets sharper as $S$ increases and becomes, in the thermodynamic limit, a nonequilibrium phase transition, in which the oscillatory regime corresponds to a time crystal \cite{Iemini2018}.
Crucially, systems of $N$ atoms with different total angular momentum $S$ undergo the crossover at  different $\omega$ values [see Fig.~\ref{fig_cartoon}(b)] and signatures of both dynamical regimes  clearly manifest in the photocounting record \cite{Cabot2023,Cabot2024}. Moreover, the photocounting  process preserves the initial total angular momentum $S$, which allows for measuring the latter without the need to reinitialize the system.

{\it Universal photocounting statistics. --} 
The central quantity of our measurement protocol is the output intensity (or time-averaged photocount): $
I_{T}(t)=\frac{1}{T}\int_{t}^{t+T} dN(\tau)$,
where $T$ is the length of the measurement time window and $dN(t)$ is a random variable that takes the value 0 when no photon is detected and 1 when a photon is detected. In the overdamped regime $\omega<\omega_\mathrm{c}(S)$, the photocounting statistics is universal (i.e. independent of $S$) and analytically known, while the transition point varies with $S$ [see Fig.~\ref{fig_cartoon}(c)]. Deviations of the counting statistics from the  universal behavior therefore allow to infer the unknown value of the total angular momentum $S$. A single realization of the photocounting  process is described by a stochastic master equation for the conditioned system state  $\hat{\mu}$ \cite{Wiseman2009}:
\begin{equation}\label{SME}
d\hat{\mu}=dN(t)\mathcal{J}\hat{\mu}+dt\big(-i \mathcal{H}+(1-\eta)\kappa \mathcal{D}[\hat{S}_-]\big) \hat{\mu} \,.
\end{equation}
The parameter $\eta\in[0,1]$ is the detection efficiency, and we have defined the superoperators $\mathcal{J}\hat{\mu}=\hat{S}_-\hat{\mu}\hat{S}_+/\langle \hat{S}_+\hat{S}_-\rangle-\hat{\mu}$ and
$\mathcal{H}\hat{\mu}=\hat{H}_\mathrm{eff}\hat{\mu}- \hat{\mu}\hat{H}^\dagger_\mathrm{eff}-(\langle \hat{H}_\mathrm{eff}\rangle-\langle\hat{H}^\dagger_\mathrm{eff}\rangle)\hat{\mu}$. The expected values are taken with respect to the conditioned state, and the effective Hamiltonian is given by:
$\hat{H}_\mathrm{eff}=\omega\hat{S}_\mathrm{x}-i\eta\frac{\kappa}{2}\hat{S}_+\hat{S}_-$.    
For a given realization the expected number of detections in the interval $[t,t+dt]$ is given by $\mathbb{E}|_{\hat{\mu}(t)}[dN(t)]=\eta \kappa \mathrm{Tr}[\hat{S}_+\hat{S}_-\hat{\mu}(t)] dt$.
Averaging over realizations we recover the master equation, i.e. $\mathbb{E}[\hat{\mu}(t)]=\hat{\rho}(t)$. Ideal photocounting corresponds to $\eta=1$, in which  Eq.~(\ref{SME}) can be replaced by a stochastic Schrödinger equation evolving pure states \cite{Wiseman2009}.

The statistics of the output intensity  can be understood using large deviations theory \cite{Garrahan2010,Paulino2024}. When $T$ is large compared to the dominant relaxation timescales of the system, the moments of $I_{T}(t)$ can be obtained from the scaled cumulant generating function in the stationary state, $\theta(s)$ (see Supplemental Material \cite{SM}). In the overdamped regime this assumes the universal form  \cite{Cabot2023,Cabot2024}:
\begin{equation}
\theta(s)=\frac{\eta\omega^2}{\kappa}(e^{-s}-1),    
\end{equation}
where $s$ is the counting field. From  the partial derivatives of $\theta(s)$ evaluated at $s=0$ we obtain the expectation value and the variance of the time-averaged  photocount \cite{SM}:
\begin{equation}\label{moments_intensity}
\mathbb{E}[I_T(t)]=\frac{\eta\omega^2}{\kappa},\quad    \mathbb{E}[I^2_T(t)] -\mathbb{E}[I_T(t)]^2=\frac{\eta\omega^2}{\kappa T}.
\end{equation}
For large $T$ 
the statistics of the output intensity tends to a Gaussian characterized by these two moments. From Eq.~(\ref{moments_intensity}) it is clear that such statistics is independent of $S$. The total angular momentum  determines instead the point at which the emission statistics deviates from Eq.~(\ref{moments_intensity}) and begins to display a $S$-dependent behavior (see e.g. Ref.~\cite{Cabot2024}).  Therefore, starting from $\omega=0$ and adiabatically ramping up the Rabi frequency, the collected signal follows the universal statistics until the system approaches  the $S$ dependent crossover. The point at which the counting statistics changes provides the estimate for $S$ [see Fig.~\ref{fig_cartoon}(c)].

{\it Adiabatic protocol. --} We focus on cases in which the ramp up timescale of $\omega(t)$ is much larger than the dominant relaxation timescales. In this case, adiabatic large deviations theory \cite{Paulino2024} shows that the statistics of the output intensity  over a time interval $\Delta t$ is controlled by the following time integral:
\begin{equation}\label{ad_scgf}
\theta_\mathrm{ad}(t,\Delta t,s)=\int_t^{t+\Delta t} d\tau\, \theta (\tau,s),    
\end{equation}
where $\theta(t,s)$ is the instantaneous scaled cumulant generating function associated with the parameters at time $t$.  In the overdamped regime, 
$\theta(t,s)=\eta\omega(t)^2(e^{-s}-1)/\kappa$.

\begin{figure}[t!]
 \centering
 \includegraphics[width=1\columnwidth]{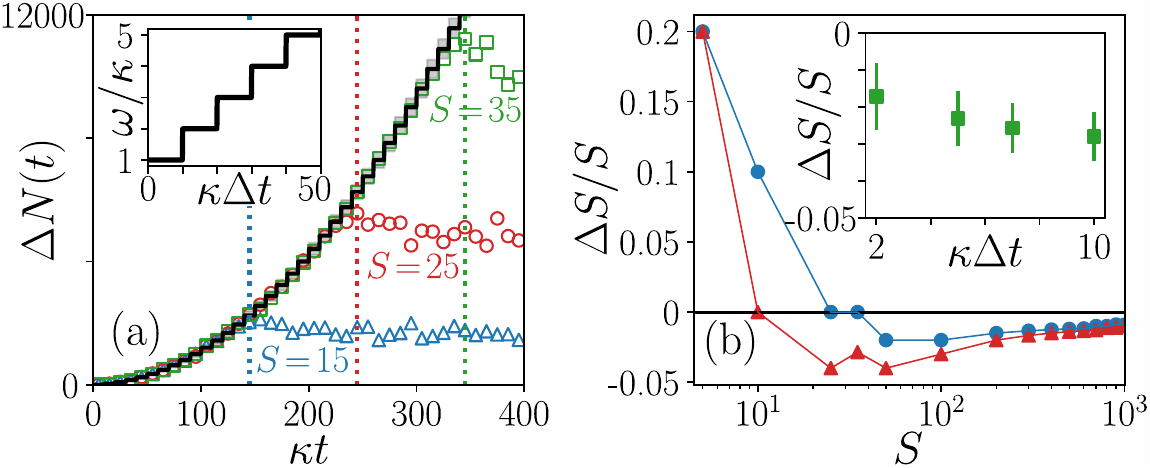}
 \caption{{\bf Adiabatic ramp up of the Rabi frequency.} (a)  Inset: adiabatic ramp given in Eq.~(\ref{rabi_ladder}). Main panel: color points correspond to single realizations of the protocol for different $S$ sectors. The black solid line corresponds to the universal curve, obtained from Eq.~(\ref{moments_intensity}), while the shadowed region corresponds to this value $\pm3$ standard deviations [see also Fig.~\ref{fig_freezing}(b),(d)]. The initial condition is fully contained in the corresponding $S$ sector, $\kappa \Delta t=10$ and $\eta=1$  ($\eta<1$ is shown in Ref.~\cite{SM}). Vertical color lines correspond to $\omega_\mathrm{c}(S)=\kappa S$, for each case. (b) Deviation of the estimated $S$ from its actual value, $\Delta S$, as a function of $S$. The deviation is computed as described in the main text, fixing $\kappa \Delta t=2$ (blue dots) or  $\kappa \Delta t=10$ (red triangles). Inset: deviation $\Delta S$ for $S=100$ and varying $\kappa \Delta t$ considering $10^3$ realizations of the measurement protocol.}
 \label{fig_adiabatic_ramp}
\end{figure}

We increase the Rabi frequency in small steps $\Delta \omega$ every $\Delta t$, where $\Delta t$ is  large compared to the relaxation timescales of the system [see Fig.~\ref{fig_adiabatic_ramp}(a)]. In this case we have a piecewise constant Rabi frequency:
\begin{equation}\label{rabi_ladder}
\omega(t)=n\Delta \omega, \,\, \,\,\text{for}\,\,\,\, (n-1)\Delta t<t\leq n\Delta t,  
\end{equation}
with $n=1,2,3\dots$ Within each time interval the number of detected photons is the stochastic variable $\Delta N(t)=\Delta t I_{\Delta t}(t)$.
The statistics of the output intensity (number of counts) is directly obtained from Eq.~(\ref{moments_intensity}) since $\theta_\mathrm{ad}(t,\Delta t,s)=\Delta t\theta(t,s)$, with the corresponding value of $\omega$. We illustrate this protocol in Fig.~\ref{fig_adiabatic_ramp}(a), considering the monitored output for different total angular momenta. The output follows the universal curve (black solid line) until  $\omega(t)=\omega_\mathrm{c}(S)$ (vertical lines), when it starts to deviate significantly from it. Such a deviation allows to infer the $S$-dependent transition point. Here, $\Delta\omega$ sets the minimum resolution on $S$ to $\Delta \omega/\kappa$, and we fix $\Delta\omega/\kappa=1$. Notice that other adiabatic protocols for $\omega(t)$ are possible, for which the universal curve would display a different behavior \cite{SM}.

The achievable precision is the result of the trade-off between two sources of error. For short time windows $\Delta t$, the standard deviation of the photocount is larger [cf.~Eq.~(\ref{moments_intensity})] resulting in higher fluctuations in individual realizations. This makes false positives, that is unexpected significant deviations from the universal curve, more common. For large time windows (and intermediate values of $S$) a systematic error appears. This is due to the fact that, for finite $S$, the system displays a crossover and not a genuine phase transition. As a consequence, the actual emission statistics deviates from the universal one slightly before  $\omega_\mathrm{c}(S)$ \cite{Cabot2023}. This is illustrated in Fig.~\ref{fig_adiabatic_ramp}(b), where we show the systematic deviation of the estimated $S$ from its actual value. The estimated $S$ is here obtained from the Rabi frequency at which the average (with respect to $\hat{\rho}$) number of detections deviates three standard deviations [computed using Eq.~(\ref{moments_intensity}) for a given $\kappa\Delta t$] from the universal curve. Except for small $S$, our protocol tends to underestimate the actual value of $S$, with short measuring time windows performing better in the limit of many realizations. Nevertheless, the relative error remains small (less than $5\%$) and  diminishes as $S$ increases. In the inset of Fig.~\ref{fig_adiabatic_ramp}(b), we illustrate the precision of our protocol for a finite number of realizations, fixing $S=100$ and considering four values of $\Delta t$. We observe that, upon increasing $\Delta t$, the systematic error increases, while the error bars are slightly larger for smaller $\Delta t$, reflecting the above-mentioned trade-off.

\begin{figure}[t!]
 \centering
 \includegraphics[width=1\columnwidth]{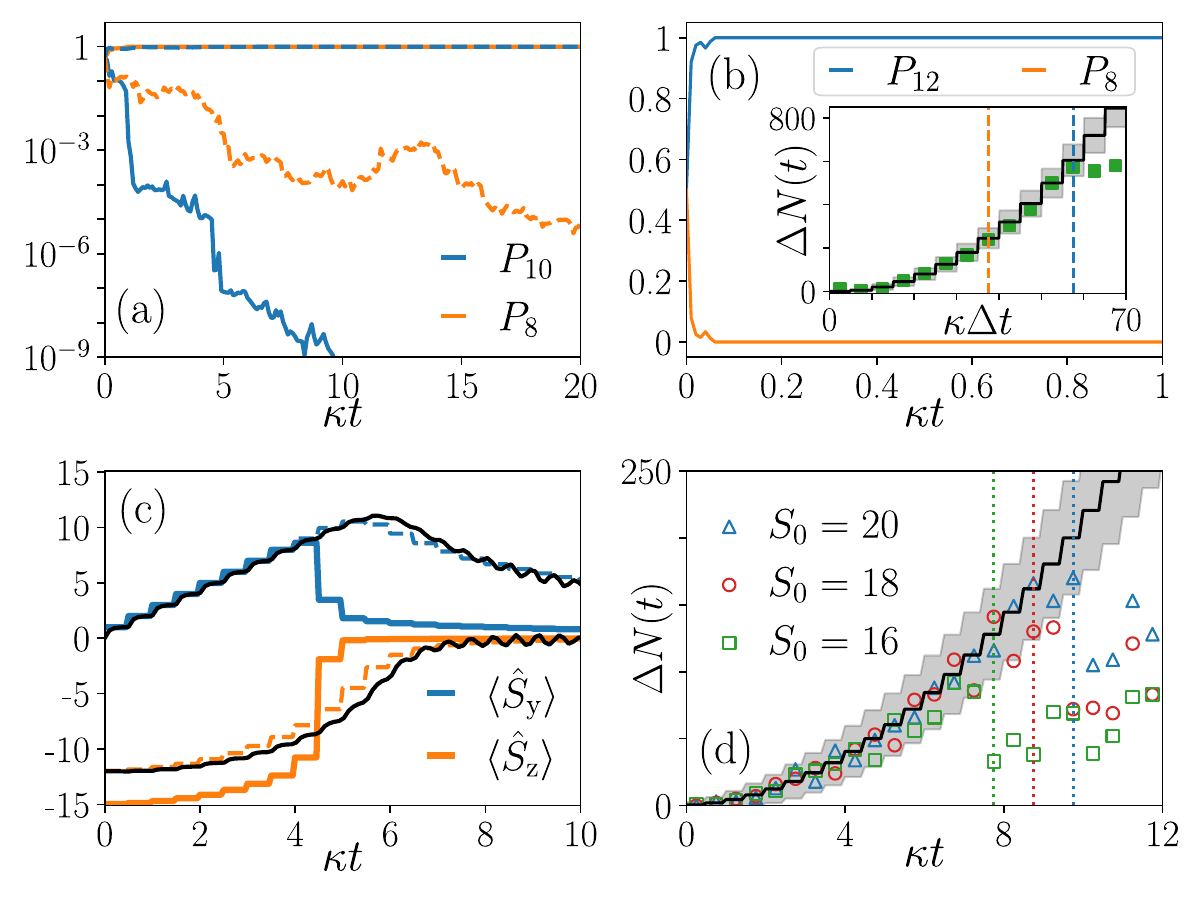}
 \caption{{\bf Dissipative freezing and local decay}. (a) Time evolution of the sector occupations $P_{S_{1,2}}$ for $S_1=10$, $S_2=8$ and $\omega/\kappa=1.5S_2$. Solid lines correspond to ideal photodetection ($\eta=1$) for the initial  superposition state $|\Psi(0)\rangle$. Dashed lines correspond to $\eta=0.5$ and the initial mixed state $\hat{\rho}(0)$. (b) Adiabatic ramping up protocol with initial state $|\Psi(0)\rangle$, $S_1=12$ and $S_2=8$ and $\kappa\Delta t=5$. Main panel: time window in which dissipative freezing takes place. Inset: number of detected photons and universal curve. The vertical lines indicate $t_\mathrm{c}(S)$ for each sector. (c-d) Impact of local decay. (c) Master equation dynamics of $\langle \hat{S}_\mathrm{y,z} \rangle$ (black solid thin lines) for the adiabatic ramp [Eq.~(\ref{rabi_ladder})] with $N=30$, $\gamma/\kappa=0.05$, $\kappa\Delta t=0.5$. The initial state is characterized by $S_0=12$ and magnetization $S_\mathrm{z}=-S_0$. Color solid (thick) lines correspond to the true stationary results for each value of $\omega(t)$, while color dashed lines to the respective values for $\gamma/\kappa=0$ and $S_0=12$. (d) Color points: photon counts for individual realizations of Eq.~(\ref{rabi_ladder}) for $N=40$, $\gamma/\kappa=0.05$, $\kappa\Delta=0.5$, $\eta=1$ and various $S_0$. Solid black line and shadow correspond to the universal curve and three standard deviations from it [Eq.~(\ref{moments_intensity})]. Vertical lines indicate the corresponding times at which $\omega(t)=\kappa S_0$.}
 \label{fig_freezing}
\end{figure}

{\it Superpositions and mixtures of total angular momentum. --} Let us now discuss the case in which the state of the system is a superposition of different total angular momentum states. In this case the measurement becomes projective, i.e. the system selects a given total angular momentum sector. This projection results from the occurrence of dissipative freezing \cite{Munoz2019,Tindall2023}. 
Dissipative freezing can be characterized considering the projectors on each symmetry sector: $\hat{P}_S=\sum_{S_\mathrm{z}=-S}^S|S,S_\mathrm{z}\rangle \langle S, S_\mathrm{z}|$, whose expected value (with respect to $\hat{\mu}$) is denoted by $P_S(t)$. By averaging over realizations $P_S(t)$ reproduces the weights of the initial state, due to conservation of total angular momentum \cite{Tindall2023}. In Fig.~\ref{fig_freezing}(a) we show the occurrence of dissipative freezing within  the oscillatory regime [$\omega>\omega_\mathrm{c}(S)$].  We show a realization for an initial superposition state $|\Psi(0)\rangle=\frac{1}{\sqrt{2}}(|S_1,0\rangle+|S_2,0\rangle)$ and $\eta=1$ (solid lines), and a realization for an initial mixed state $\hat{\rho}(0)=\frac{1}{2}(|S_1,0\rangle\langle S_1,0|+|S_2,0\rangle\langle S_2,0|)$ with finite detection efficiency $\eta=0.5$ (dashed lines). In the overdamped regime [$\omega<\omega_\mathrm{c}(S)$], dissipative freezing occurs only partially as the system is able to reach a stationary state (within a single realization) that still spans more than one symmetry sector \cite{SM}. The weights on the different sectors vary during the transient to this stationary state, and  their final value  depends on the realization. Nevertheless, this has no consequences for our protocol, as complete projection eventually occurs when the Rabi frequency becomes close to the smaller $\omega_\mathrm{c}(S)$ among the involved sectors.

In Fig.~\ref{fig_freezing}(b) we illustrate the application of the adiabatic protocol in Eq.~(\ref{rabi_ladder}) to the initial superposition state $|\Psi(0)\rangle$.  The main panel displays the evolution of $P_S(t)$ for a realization of the protocol considering an intermediate measuring time window $\kappa\Delta t=5$. The projection into one of the sectors occurs on a timescale that is much shorter than the duration of the protocol. Thus, after a fast transient, the protocol faces essentially the scenario analyzed in Fig.~\ref{fig_adiabatic_ramp}. The inset shows the photocounts of the corresponding realization. The signal deviates from the universal curve at the point corresponding to the $S$ sector to which the system has collapsed.

{\it Local decay. --} We now address the effects of local decay on the protocol. This process introduces $N$ additional jump operators and the system is described by:
\begin{equation}\label{ME_model2}
\hat{H}=\omega\hat{S}_\mathrm{x},\quad \hat{L}=\sqrt{\kappa} \hat{S}_-,\quad \hat{L}_j=\sqrt{\gamma}\hat{\sigma}_-^{(j)}\,\,\,\,\forall j.
\end{equation}
We assume initial states to have equal weights on all irreducible representations with total angular momentum $S_0$ and same magnetization $S_\mathrm{z}$ in each of these. Following Refs.~\cite{Chase2008,Baragiola2010}, the dynamics of such states can be efficiently investigated even if, for $\gamma>0$, the total angular momentum is not conserved. This latter aspect also implies the existence of a unique stationary state, generally contained in more than one $S$ sector.  Nevertheless, the manifold of stationary states (one for each value of $S$) of Eq.~(\ref{ME}) and (\ref{ME_model1}) still manifests, although in a metastable fashion, when $\gamma/\kappa$ is small \cite{SM}. This allows us to estimate the total angular momentum of the initial state as before. A crucial requirement is however to perform the protocol as fast as possible in order to minimise the detrimental effects of local decay. 

In Fig.~\ref{fig_freezing}(c) we illustrate the emergent metastable dynamics for $N=30$ and $\gamma/\kappa=0.05$. The system is initialized in a state with total angular momentum $S_0=12$ and subject to the adiabatic protocol given in Eq.~(\ref{rabi_ladder}). The expectations $\langle \hat{S}_\mathrm{y,z} (t)\rangle$ (black solid lines) follow closely the values corresponding to the stationary state for $S_0$ and $\gamma/\kappa=0$ (dashed lines) instead of the ones of the true stationary state (color solid lines). This metastable behavior is more robust for larger  $S_0$  at fixed $N$ \cite{SM}. Metastability is controlled by the leading eigenmodes of  the Liouvillian, that is, those associated with a smaller decay rate \cite{Macieszczak2016b,Rose2016,Cabot2021,Cabot2022}. For small $\gamma/\kappa$ there is a set of $N/2$ eigenmodes with a decay rate much smaller than the rest. Together with the stationary state, these define a manifold whose properties reflect those of the stationary states for $\gamma/\kappa=0$ \cite{SM}.  As the strength of the local losses is increased, the decay rate of these eigenmodes also increases, some of them being more affected than the others, making the sectors with smaller $S_0$ more susceptible to local decay.

We now consider the situation in which the collective emission channel is monitored. This scenario emerges naturally when collective emission is induced by, e.g., an optical cavity or waveguide [Fig.~\ref{fig_cartoon}(a)]. The photocounting process is then described by a stochastic master equation that can be simulated efficiently using a permutation invariant representation \cite{Bargiola2010,Shammah2018,SM}. In Fig.~\ref{fig_freezing}(d) we show single realizations of the protocol for $N=40$, $\gamma/\kappa=0.05$ and different values of $S_0$, taking a short measurement window $\kappa\Delta t=0.5$.  The lower bound on $\kappa\Delta t$ is dictated by the validity of the adiabatic approximation [Eq.~(\ref{ad_scgf})]. In this sense, we need to be adiabatic with respect to the relaxation onto the metastable manifold and not with respect to the ultimate relaxation to the actual stationary state. The main drawback of reducing $\kappa\Delta t$ are the stronger fluctuations in the photocount [Eq.~(\ref{moments_intensity})], which lead to larger errors when estimating the total angular momentum with a single realization. The behavior observed in Fig.~\ref{fig_freezing}(d) is qualitatively similar to the case with $\gamma/\kappa=0$, and the individual trajectories deviate from the universal curve where expected. As shown in Ref.~\cite{SM},  trajectories corresponding to different $S_0$ remain distinguishable up to $\gamma/\kappa\sim 0.1$, at least for the largest half of possible $S_0$ values. For  $\gamma/\kappa> 0$,  trajectories tend to deviate from the universal curve later than they should (except for $S_0=N/2$). The estimation protocol might here benefit from recalibrating $\omega_\mathrm{c}(S)$ using the exact dynamics.

{\it Conclusions. --} We have shown that the combination of strong symmetries and nonequilibrium phase transitions can be exploited for sensing applications. By engineering collective spin systems, one can make use of these resources to estimate a collective property of the spin ensemble (total angular momentum) through continuous monitoring. Since such a quantity is conserved by the dissipative dynamics, the measurement protocol can be iterated without the need of reinitializing the state of the ensemble. In the case of initial states without a well defined symmetry, the protocol performs a projective measurement through the action of dissipative freezing. The main difficulty faced by the proposed protocol is posed by local decay, which needs to be sufficiently weak such that the symmetry is conserved at a metastable level. In this sense, it would be interesting to explore whether the effects of local decay can be minimized by other adiabatic protocols. It would also be intriguing to apply these ideas to other many-body scenarios that combine strong symmetries with nonequilibrium transitions.

\begin{acknowledgements}
{\it Acknowledgements. --} We are grateful for financing from the Baden-Württemberg Stiftung through Project No. BWSTISF2019-23. AC acknowledges support from the Deutsche Forschungsgemeinschaft (DFG, German Research Foundation) through the Walter Benjamin programme, Grant No. 519847240. FC~is indebted to the Baden-W\"urttemberg Stiftung for the financial support of this research project by the Eliteprogramme for Postdocs. We acknowledge the use of Qutip python library \cite{Qutip1,Qutip2}. We acknowledge funding from the Deutsche Forschungsgemeinschaft (DFG, German Research Foundation) through the Research Unit FOR 5413/1, Grant No.~465199066. We acknowledge support by the state of Baden-Württemberg through bwHPC and the German Research Foundation (DFG) through grant no INST 40/575-1 FUGG (JUSTUS 2 cluster). This work was developed within the QuantERA II Programme (project CoQuaDis, DFG Grant No. 532763411) that has received funding from the EU H2020 research and innovation programme under GA No. 101017733.
\end{acknowledgements}

\bibliographystyle{apsrev4-2}
\bibliography{references}


\setcounter{equation}{0}
\setcounter{figure}{0}
\setcounter{table}{0}
\makeatletter
\renewcommand{\theequation}{S\arabic{equation}}
\renewcommand{\thefigure}{S\arabic{figure}}

\makeatletter
\renewcommand{\theequation}{S\arabic{equation}}
\renewcommand{\thefigure}{S\arabic{figure}}

\onecolumngrid
\newpage

\setcounter{page}{1}

\begin{center}
{\Large SUPPLEMENTAL MATERIAL}
\end{center}
\begin{center}
\vspace{1.3cm}
{\Large Exploiting nonequilibrium phase transitions and strong symmetries for continuous measurement of collective observables
}
\end{center}
\begin{center}
Albert Cabot$^{1}$, Federico Carollo$^{1}$, Igor Lesanovsky$^{1,2,3}$
\end{center}
\begin{center}
$^1${\it Institut f\"ur Theoretische Physik, Universit\"at T\"ubingen,}\\
{\it Auf der Morgenstelle 14, 72076 T\"ubingen, Germany}\\
$^2${\it School of Physics, Astronomy, University of Nottingham, Nottingham, NG7 2RD, UK.}\\
$^3${\it Centre for the Mathematics, Theoretical Physics of Quantum Non-Equilibrium Systems,
University of Nottingham, Nottingham, NG7 2RD, UK}
\end{center}
\section{Overview of large deviations for open quantum systems}

In the large deviation approach to open quantum systems, the central quantity is given by the partition function of the time-integrated photocount \cite{Garrahan2010}:
\begin{equation}
Z_T(s)=\mathbb{E}\big[e^{-sT I_T}\big].
\end{equation}
This quantity contains the information of all moments and cumulants of $I_T$, which can be retrieved taking partial derivatives  with respect to the counting field $s$, around $s=0$. In particular, the mean and the variance read: 
\begin{equation}
\mathbb{E}[I_T]=-\frac{1}{T}\partial_s Z_T(s)\big|_{s=0}, \quad
\mathbb{E}[I^2_T]-\mathbb{E}[I_T]^2=\frac{1}{T^2}\partial^2_s \log [Z_T(s)]\big|_{s=0}.  
\end{equation}
The partition function can be calculated as $Z_T(s)=\text{Tr}[\hat{\rho}_s(T)]$, where $\hat{\rho}_s(T)$ is the solution of a tilted master equation \cite{Garrahan2010,Carollo2018}. In the case of the counting process described by the stochastic master equation (\ref{SME}), the corresponding tilted master equation is given by: 
\begin{equation}\label{tilted_ME}
\frac{d}{dT} \hat{\rho}_s=-i[\omega \hat{S}_\mathrm{x},\hat{\rho}_s]+\kappa\mathcal{D}[\hat{S}_-]\hat{\rho}_s+\eta\kappa(e^{-s}-1)\hat{S}_-\hat{\rho}_s\hat{S}_+. 
\end{equation}
From this equation we can define the tilted Liouvillian $\partial_T\hat{\rho}_s=\mathcal{L}(s)\hat{\rho}_s$, which preserves the positivity of $\hat{\rho}_s$ but not its trace. Its dominant eigenvalue, $\theta(s)$, is real and generally different from zero. This corresponds to the (stationary) {\it scaled cumulant generating function}. All cumulants can be obtained from $\theta(s)$ using its relation with the partition function:
\begin{equation}\label{SCGF}
\lim_{T\to\infty}\frac{1}{T}\log[Z_T(s)]= \theta(s).   
\end{equation}
Therefore, for large measurement times $T$, we have that:
\begin{equation}
\lim_{T\to\infty}\mathbb{E}[I_T]=-\partial_s \theta(s)|_{s=0},  \quad \lim_{T\to\infty} \mathbb{E}[I^2_T]-\mathbb{E}[I_T]^2=\frac{1}{T}\partial^2_s \theta(s)|_{s=0}.   
\end{equation}

\section{Supplemental results for  the adiabatic ramping up protocol}

\begin{figure}[h!]
 \centering
 \includegraphics[width=0.6\columnwidth]{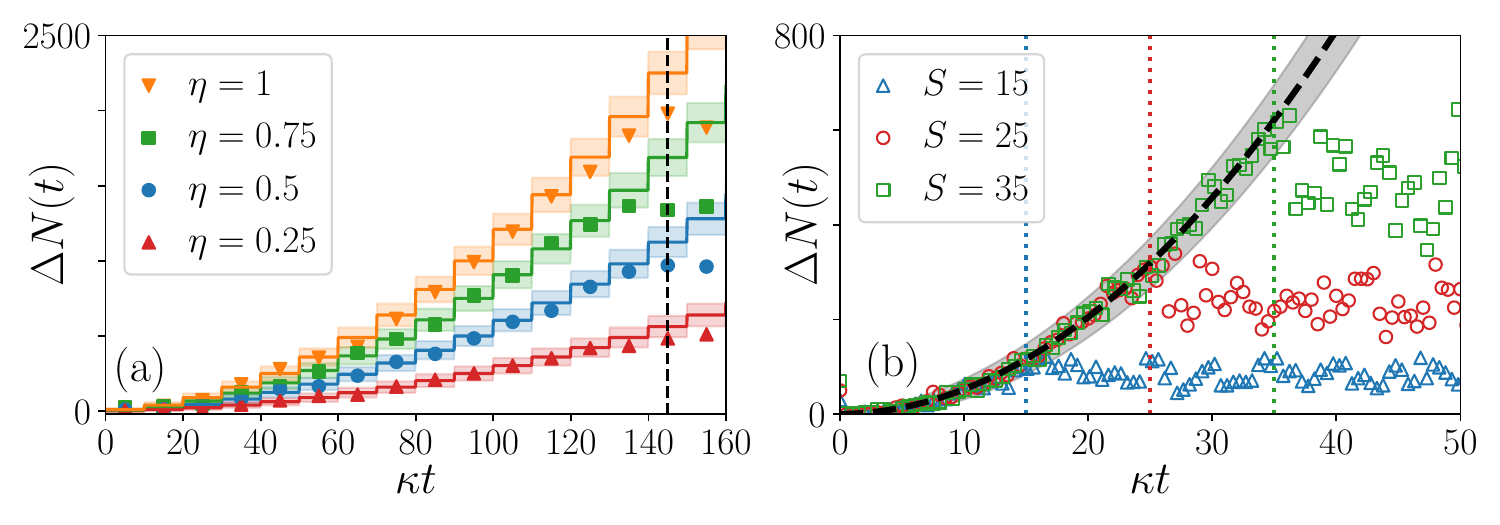}
 \caption{{\bf Supplemental results for  the adiabatic ramping up protocol.} (a) Effects of decreasing the detection efficiency $\eta$ for the ramp up protocol given in Eq.~(\ref{rabi_ladder}). Color points: results for a single realization of the protocol with $\kappa\Delta t=10$ and $S=30$. Color solid lines: expected value of detections according to the universal curve $\pm 3$ standard deviations.  (b) Alternative measurement protocol for the adiabatic ramp given in Eq.~(\ref{rabi_linear}) for $\kappa \Delta t=0.5$ and $\alpha/\kappa^2=1$. Black dashed line corresponds to the expected number of detections [computed from Eq.~(\ref{linear_intensity})], while shadowed region corresponds to this value $\pm3$ standard deviations. We consider an initial condition fully contained the corresponding $S$ sector and $\eta=1$. Vertical color lines correspond to $\omega_\mathrm{c}(S)=\kappa S$ for each case.}
 \label{fig_SM_alternative protocol}
\end{figure}

\subsection{Finite detection efficiency}

In Fig.~\ref{fig_SM_alternative protocol}(a) we illustrate the effects of a finite detection efficiency $\eta<1$ for the adiabatic protocol discussed in the main text [Eq.~(\ref{rabi_ladder})]. We consider a system initialized in the total angular momentum $S=30$ and the protocol is realized with $\kappa\Delta t=10$. For all the considered values of $\eta$, we observe the points to follow the corresponding universal curve until $\omega(t)=\omega_\mathrm{c}(S)$. As expected from Eq.~(\ref{moments_intensity}), the effect of decreasing the detection efficiency is to reduce the number of detected photons by the factor $\eta$.

\subsection{Alternative ramping up protocol}

In this section we consider a different adiabatic protocol for ramping up the Rabi frequency that leads to a different universal curve. In particular we consider a Rabi frequency that increases linearly in time as:
\begin{equation}\label{rabi_linear}
\omega(t)=\alpha t.    
\end{equation}
Using adiabatic large deviations theory [Eq.~(\ref{ad_scgf})] and
$\theta(t,s)=\eta\omega(t)^2(e^{-s}-1)/\kappa$, we obtain:
\begin{equation}
\theta_\mathrm{ad}(t,\Delta t,s)=\frac{\alpha^2\eta}{3\kappa}(3t^2\Delta t+3t\Delta t^2+\Delta t^3)(e^{-s}-1),     
\end{equation}
which describes the emission statistics for the linear adiabatic ramp within  the overdamped regime. From this expression we can compute the first two moments of the output intensity for the interval $\Delta t$:
\begin{equation}\label{linear_intensity}
\text{E}[I_{\Delta t}(t)]=\frac{\eta\alpha^2}{3\kappa}(3t^2+3t\Delta t+\Delta t^2),\quad \text{E}[I_{\Delta t}^2(t)]-\text{E}[I_{\Delta t}(t)]^2=\frac{1}{\Delta t}\text{E}[I_{\Delta t}(t)].
\end{equation}
Analogously to the case presented in the main text, the protocol  consists in increasing the Rabi frequency such as Eq.~(\ref{rabi_linear}) and measuring the detected photons in intervals $\Delta t$, such that at each time interval the number of detected photons is the stochastic variable $\Delta N(t)=\Delta t I_{\Delta t}(t)$. From the behavior of $\Delta N(t)$ we infer when the system has crossed the point $\omega_\mathrm{c}(S)=\kappa S$, and thus $S$. In the linear adiabatic ramp, the parameters $\alpha$ and $\Delta t$ influence the resolution we have on $S$ as $\Delta S=\alpha \Delta t/\kappa$, which we want to keep below one. We show this adiabatic ramp protocol in Fig.~\ref{fig_SM_alternative protocol}(b), for three values of $S$ and an initial condition residing entirely in each corresponding total angular momentum sector. We observe the counts to follow the universal curve, while they begin to deviate from it around each of the corresponding $\omega_\mathrm{c}(S)$ (vertical color lines).

\section{Analysis of the effects of local decay}

This section presents some supplemental results about the effects of local decay on the adiabatic protocol presented in the main text. For this, we study the dynamics at the master equation level and also for quantum trajectories (in which only collective emissions are monitored). We also analyze the signatures of metastability in the Liouvillian spectrum. In this study, we make use of permutation invariant representations \cite{Chase2008,Baragiola2010,Shammah2018} for the operators of the system and the considered class of initial states. 

\subsection{Permutation invariant quantum jump trajectories}

\begin{figure}[t!]
 \centering
 \includegraphics[width=0.7\columnwidth]{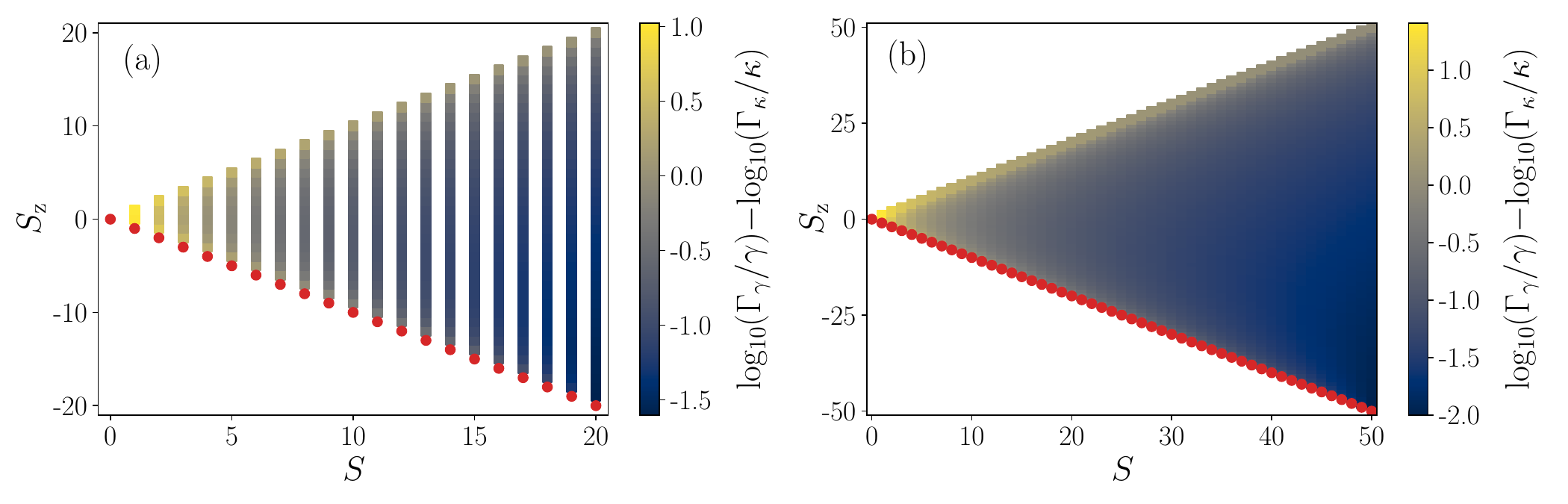}
 \caption{{\bf Ratio of rate of local jumps with respect to collective ones for the total angular momentum states.} (a) $N=40$, (b) $N=100$. In color we have the ratio in base 10 logarithm. Notice that we plot $\kappa \Gamma_\gamma/(\gamma \Gamma_\kappa)$ and thus we factor out the ratio $\gamma/\kappa$. The red dots correspond to the points in which $\Gamma_\kappa(N,S)=0$, for which the ratio is not well defined.}
 \label{fig_SM_ratio}
\end{figure}

In the presence of local decay with rate $\gamma$ the master equation is modified to:
\begin{equation}\label{PI_ME}
\partial_t \hat{\rho}=-i[\omega \hat{S}_\mathrm{x},\hat{\rho}]+\kappa\big(\hat{S}_- \hat{\rho}\hat{S}_+-\frac{1}{2}\{\hat{S}_+ \hat{S}_-, \hat{\rho}\}\big)+\gamma\sum_{j=1}^N\big(\hat{\sigma}^{(j)}_- \hat{\rho}\hat{\sigma}^{(j)}_+-\frac{1}{2}\{\hat{\sigma}^{(j)}_+ \hat{\sigma}^{(j)}_-, \hat{\rho}\}\big).
\end{equation}
This equation no longer conserves total angular momentum. However, it still has a permutation symmetry, as it is invariant under permutation of the spin labels. This allows one to efficiently simulate it  by means of a permutation invariant representation of the state and operators \cite{Bargiola2010,Shammah2018}. In this sense, it is also necessary to restrict the study to states that are identical with respect to the degenerate representations of total angular momentum for a given $S$ and $N$ \cite{Chase2008,Bargiola2010}.

When considering the unravelling of Eq.~(\ref{PI_ME}) in quantum jump trajectories, one has to take into account that quantum jumps associated to a specific spin, $\hat{L}_j=\sqrt{\gamma}\hat{\sigma}_-^{(j)}$, break the permutation symmetry of the dynamics. Therefore, the efficient permutation invariant representations can only be used if jumps are monitored in a permutation invariant way, in which there is no knowledge about which spin has emitted which photon. This prevents the use of a stochastic Schrodinger equation unravelling for pure states, as it requires to specify this knowledge. Nevertheless, a permutation invariant unravelling can be written down in terms of a stochastic master equation, in which the stochastic term is implemented by the permutation invariant representation of local decay.

In this work we focus on a simpler scenario in which only collective emissions are monitored, while local ones are not. This situation emerges naturally when collective emission is induced by, e.g., an optical cavity and only the output of the cavity is monitored. The stochastic master equation describing this case is also suited to the efficient permutation invariant representation of Refs. \cite{Chase2008,Baragiola2010,Shammah2018}. This can be written as:
\begin{equation}\label{PI_SME}
d\hat{\mu}=dN(t)\mathcal{J}\hat{\mu}+dt\bigg(-i \mathcal{H}+(1-\eta)\kappa \mathcal{D}[\hat{S}_-]+\gamma\sum_{j=1}^N \mathcal{D}[\hat{\sigma}^{(j)}_-]\bigg) \hat{\mu}, 
\end{equation}
with
\begin{equation}
\mathcal{J}\hat{\mu}=\frac{\hat{S}_-\hat{\mu}\hat{S}_+}{\langle \hat{S}_+\hat{S}_-\rangle}-\hat{\mu},\quad \mathcal{H}\hat{\mu}=\hat{H}_\mathrm{eff}\hat{\mu}- \hat{\mu}\hat{H}^\dagger_\mathrm{eff}-(\langle \hat{H}_\mathrm{eff}\rangle-\langle\hat{H}^\dagger_\mathrm{eff}\rangle)\hat{\mu},\quad \hat{H}_\mathrm{eff}=\omega\hat{S}_\mathrm{x}-i\eta\frac{\kappa}{2}\hat{S}_+\hat{S}_-.    
\end{equation}
$dN(t)$ is a random variable that takes the value 0 when no (collectively emitted) photon is detected and 1 when a photon is detected, $\eta\in[0,1]$ is the detection efficiency of the collective emission channel and expectation values $\langle \dots \rangle$ are taken with respect to the conditioned state $\hat{\mu}$. For a given realization the expected number of photodetections at the interval $[t,t+dt]$ is given by: $\mathbb{E}|_{\hat{\mu}(t)}[dN(t)]=\eta \kappa \mathrm{Tr}[\hat{S}_+\hat{S}_-\hat{\mu}(t)] dt$. 
\begin{figure}[t!]
 \centering
 \includegraphics[width=0.8\columnwidth]{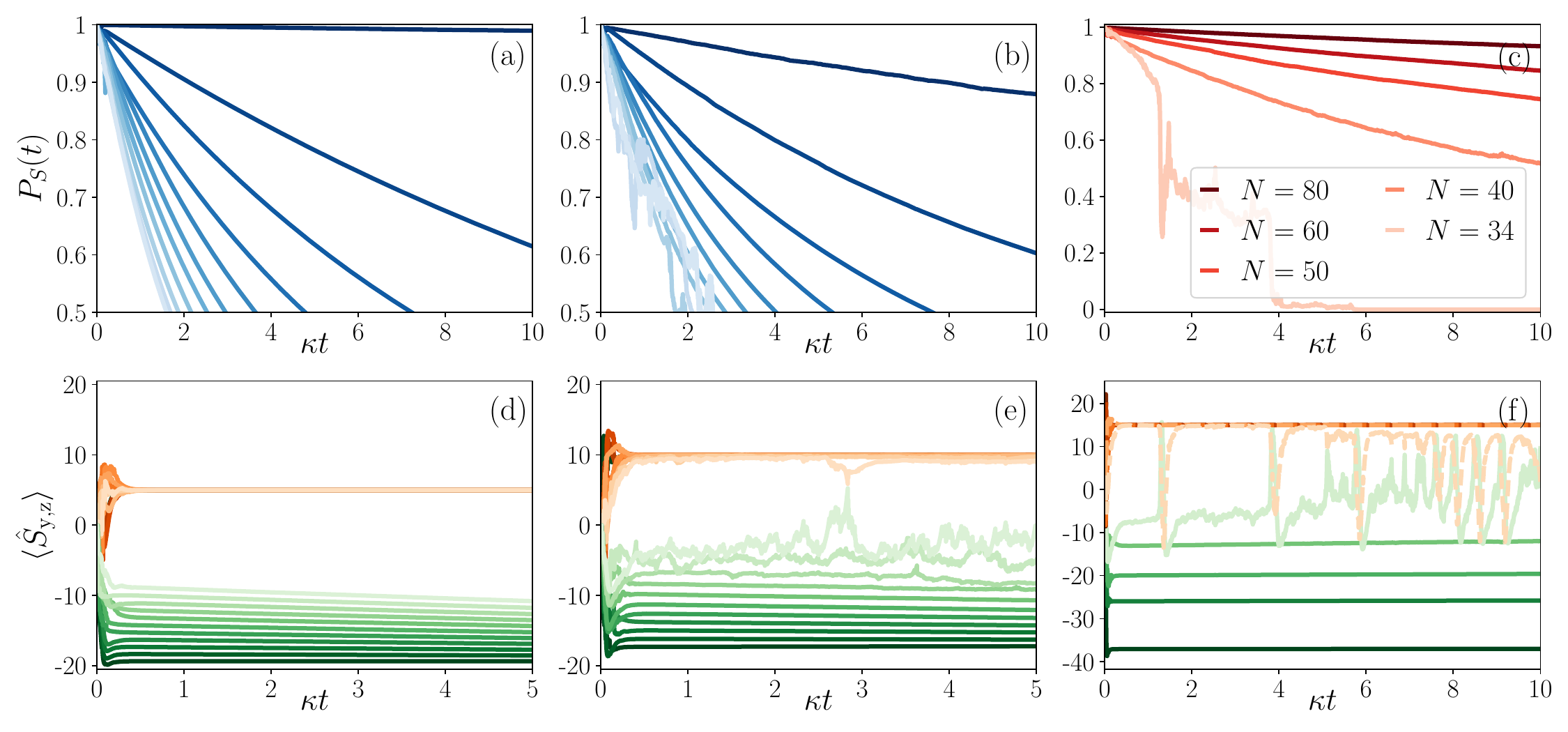}
 \caption{{\bf Decay of the total angular momentum.} (a) Weight in the $S$ total angular momentum sector, $P_S(t)$, for an initial state characterized by a total angular momentum $S$ and a z-magnetization $S_\mathrm{z}=0$ for a single realization of Eq.~(\ref{PI_SME}). Each line corresponds to a different $S$ (both in $P_S(t)$ and in the initial state), with $S\in[10,20]$ and larger $S$ being represented by darker colors. Parameters: $N=40$, $\omega/\kappa=0.125N$, $\eta=1$ and $\gamma/\kappa=0.05$. (b) Same as in (a) but for $\omega/\kappa=0.25N$. (c) Weight in the $S$ total angular momentum sector, $P_S(t)$, for the same kind of initial state  as before  and a single realization of Eq.~(\ref{PI_SME}), considering $S=N/2$ and varying $N$. Parameters: $\omega/\kappa=15$, $\eta=1$ and $\gamma/\kappa=0.05$. (d)-(f) Collective magnetizations for the trajectories corresponding to panels (a)-(c), respectively. Green colors correspond to $\langle \hat{S}_\mathrm{z}\rangle$, while orange colors to $\langle \hat{S}_\mathrm{y}\rangle$.}
 \label{fig_SM_decay}
\end{figure}

\subsection{Decay of total angular momentum for different $S$ and $N$}

We first analyze the relaxation timescales of total angular momentum as parameters are varied. For this, we consider initial conditions contained entirely in one of the sectors $S$ and study the timescales characterizing the spreading of the state into other $S$ sectors as the initial $S$, $N$, $\omega/\kappa$ and $\gamma/\kappa$ are varied.  

Before analyzing the dynamics of the system, it is illustrative to consider how the rate of collective jumps for a given total angular momentum state and the rate of local jumps for the same state vary with $S$, $S_\mathrm{z}$ and $N$. These quantities are defined as:
\begin{equation}
\begin{split}
&\Gamma_\kappa(N,S)=\kappa \langle S,S_\mathrm{z}|\hat{S}_+\hat{S}_-| S,S_\mathrm{z}\rangle= \kappa \big[S(S+1)-S_\mathrm{z}(S_\mathrm{z}-1)\big],\\
&\Gamma_\gamma(N,S)=\gamma\langle S,S_\mathrm{z}|\sum_{j=1}^N \hat{\sigma}^{(j)}_+ \hat{\sigma}^{(j)}_-| S,S_\mathrm{z}\rangle= \frac{\gamma}{2} \big(N+2S_\mathrm{z}).
\end{split}
\end{equation}
The ratio between the rate of local jumps and that of collective jumps is plotted in Fig.~\ref{fig_SM_ratio} for $N=40$ (a) and $N=100$ (b), for all possible choices of $S$ and $S_\mathrm{z}$. The bigger the ratio, the more important is the effect of the local decay channels. We observe that the sector with maximal $S$ is in general the more robust against local decay channels, while intermediate sectors are more susceptible to them. Close to the maximal $S$ sectors, larger $N$ also makes the rate of local jumps smaller with respect to the collective one. The value of $S_\mathrm{z}$ plays an important role too, and states with negative z-magnetization are in general less susceptible to local decay. In this figure, we have factored out the constant $\gamma/\kappa$, as it just provides a constant scale factor that can favor overall one type of channel over the other. 

In Fig.~\ref{fig_SM_decay} we consider the dynamics of Eq.~(\ref{PI_SME}) for initial states contained in just one $S$ sector. We analyze how the weight in the initial sector progressively decays out as the system explores other $S$ sectors. In Fig.~\ref{fig_SM_decay}(a-b) we fix $N=40$ and we vary the initial value of $S$. We observe that the closer $S$ is to its maximum value, the more time the state of the system remains in that initial sector. Comparing (a) and (b) we also observe that the value of $\omega/\kappa$ influences this spreading timescale. In general, these results cannot be simply understood from the interplay of just $\Gamma_\gamma(N,S)$ and $\Gamma_\kappa(N,S)$. However, these make the correct qualitative prediction that for intermediate values of $S$ the effects of local decay are more important. In Fig.~\ref{fig_SM_decay}(c), we fix $\omega/\kappa$, $S=N/2$ and we increase $N$, finding that the larger is the system size, the slower is the spreading from the maximal $S$ to the other sectors. In Fig.~\ref{fig_SM_decay}(d) to (f) we show the corresponding $y$ and $z$ magnetizations for the same trajectories of panels (a) to (c). We observe that the effects of local decay are not so apparent as in the projectors $P_S(t)$. This is analyzed in more detail in the next section.

\subsection{Metastable adiabatic protocol for different $S$ and $N$}

In the following we analyze how the effects of local decay manifest in the adiabatic ramp protocol studied in the main text [Eq.~(\ref{rabi_ladder})]. 

\begin{figure}[t!]
 \centering
 \includegraphics[width=1\columnwidth]{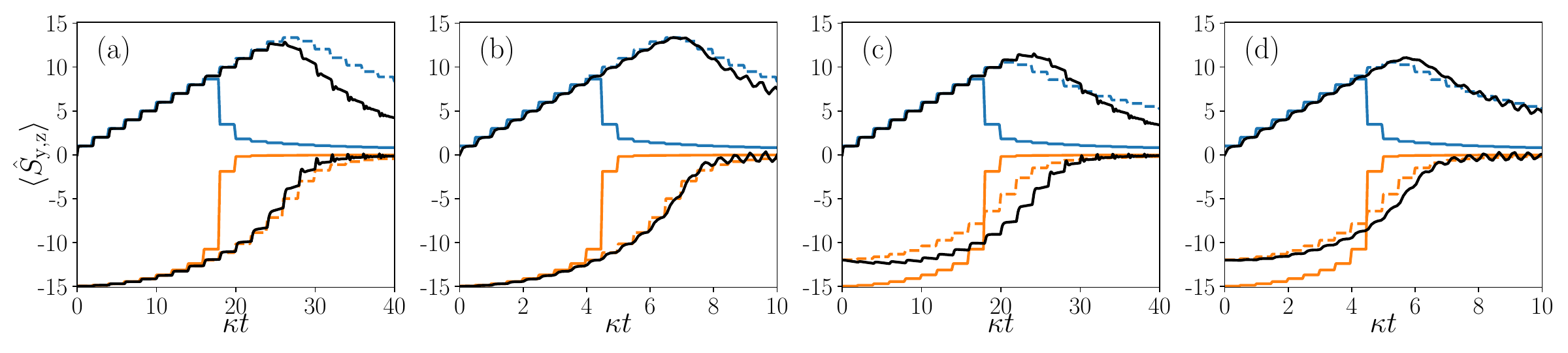}
 \caption{{\bf Metastability in the presence of local losses for $N=30$.} In all panels the adiabatic ramp up protocol given by Eq.~(\ref{rabi_ladder}) is performed in the presence of local losses with $\gamma/\kappa=0.05$. Solid color lines represent the true stationary state (for each value of $\omega(t)$), while dashed lines correspond to the stationary state for the initially considered $S$ and $\gamma/\kappa=0$. We use the color blue (orange) for the $y$ ($z$) component of the collective magnetization. Black solid lines correspond to the dynamics of $\langle \hat{S}_\mathrm{x,y}\rangle$ for the adiabatic ramp  with $\gamma/\kappa=0.05$, and initial condition with total angular momentum $S_0$ and magnetization $S_z=-S_0$. Parameters: (a) $S_0=15$ and $\kappa\Delta t=2$; (b) $S_0=15$ and $\kappa\Delta t=0.5$; (c) $S_0=12$ and $\kappa\Delta t=2$; (d) $S_0=12$ and $\kappa\Delta t=0.5$. In all cases $\eta=1$.}
 \label{fig_SM_meta_1}
\end{figure}

In Fig. \ref{fig_SM_meta_1} we consider a system of $N=30$ particles with an initial state fully contained in a sector of total angular momentum $S_0$ and with $\gamma/\kappa=0.05$. We then apply the discrete adiabatic ramp of the Rabi frequency [Eq.~(\ref{rabi_ladder})], considering different initial total angular momentums $S_0$ and protocol step lengths $\kappa\Delta t$. We study the system at the master equation level, and we compare the results of the dynamics (black solid lines) with the true stationary state for each value of $\omega(t)$ (color solid lines) and the corresponding stationary stationary state for $\gamma/\kappa=0$ (color dashed lines). Considering the case $S_0=N/2$ [panels (a) and (b)], we observe that the dynamics follows quite closely the corresponding stationary state for $\gamma/\kappa=0$. Deviations from this curve are only observed for long times. In fact, reducing the protocol step $\kappa\Delta t$, makes these deviations less significant. However, the drawback of reducing  $\kappa\Delta t$ is that the dynamics is less adiabatic and in single realizations of the photocount process fluctuations are larger. Considering $S_0=12$ [panels (c) and (d)], we observe that deviations from the curves corresponding to $\gamma/\kappa=0$ are more significant. Nevertheless, the dynamics takes closer values to the case with $\gamma/\kappa=0$ (color dashed lines) than to the case with  $\gamma/\kappa=0.05$ (color solid lines). Actually, for  $\kappa\Delta t=0.5$ [panel (d)] deviations from the dashed lines are quite small. The results of Fig.~\ref{fig_SM_meta_1} point out that, for small $\gamma/\kappa$, the system displays a metastable dynamics that is quite close to the case without local decay.

\begin{figure}[t!]
 \centering
 \includegraphics[width=1\columnwidth]{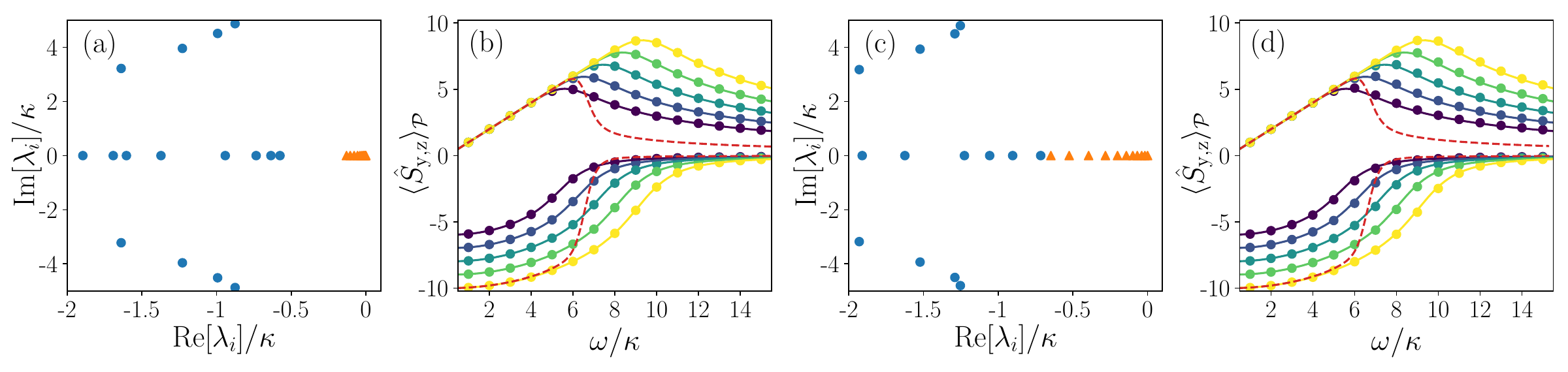}
 \caption{{\bf Metastable manifold in the Liouvillian for $N=20$.} (a) Leading Liouvillian eigenvalues for $\gamma/\kappa=0.01$ and $\omega/\kappa=5$. The stationary state together with the first $N/2$ eigenvalues with negative real part are plotted as orange triangles. (b) Color points: expected value of the $y$ and $z$ components in the metastable manifold, $\langle \hat{S}_\mathrm{y,z}\rangle_\mathcal{P}=\text{Tr}[\hat{S}_\mathrm{y,z}\mathcal{P}\hat{\rho}_{S_0}]$, where $\hat{\rho}_{S_0}$ is a state fully contained in the $S_0$ sector with zero magnetization in $S_\mathrm{z}$. The projection $\mathcal{P}$ is defined in Eq.~(\ref{projected_manifold}). The color indicates the value of $S_0$ of the  state $\hat{\rho}_{S_0}$, with $S_0\in\{6,7,8,9,10\}$ (the smaller $S_0$ the darker the color). Solid lines: expected value of the $y$ and $z$ components in the stationary state for $\gamma/\kappa=0$, where the color indicates the corresponding total angular momentum sector. Red dashed line: expected value of the $y$ and $z$ components in the true stationary state of the system with local losses. (c) Same as (a) but for $\gamma/\kappa=0.05$. (d) Same as (b) but for $\gamma/\kappa=0.05$, and in the case of $\langle \hat{S}_\mathrm{y,z}\rangle_\mathcal{P}$ considering the z-magnetization of $\hat{\rho}_{S_0}$ to be $-S_0$ for the projection into the metastable manifold.}
 \label{fig_SM_meta_2}
\end{figure}

In Fig.~\ref{fig_SM_meta_2} we analyze in more detail the origin of this metastable response. For this we consider the spectral decomposition of the Liouvillian in terms of its eigenvalues and right and left eigenmatrices:
\begin{equation}
\mathcal{L}\hat{r}_j=\lambda_j\hat{r}_j, \quad \hat{l}_j^\dagger \mathcal{L}=\lambda_j\hat{l}_j^\dagger,\quad \text{Tr}[\hat{l}_j^\dagger \hat{r}_k]=\delta_{jk}.    
\end{equation}
The eigenvalues are nonpositive and are ordered such that Re$[\lambda_j]\geq$Re$[\lambda_{j+1}]$. Moreover, we have that $\lambda_0=0$, $\hat{r}_0=\hat{\rho}_\mathrm{ss}$, and $\hat{l}_0$ is the identity. The leading eigenvalues of the Liouvillian are shown in Fig.~\ref{fig_SM_meta_2} for $N=20$, $\omega/\kappa=5$ and $\gamma/\kappa=0.01$ [panel (a)] or $\gamma/\kappa=0.05$ [panel (c)]. The zero eigenvalue together with the subsequent $N/2$ ones are shown as orange triangles. For $\gamma/\kappa=0.01$ there is a clear spectral gap between this set of eigenvalues and the rest. For $\gamma/\kappa=0.05$ a significant gap is only maintained for the smallest ones of this set. This spectral gap is a well known signature of metastability \cite{Macieszczak2016}, i.e. a pre-stationary regime characterised by a slow relaxation dynamics. This is better understood when considering the spectral decomposition of the dynamics of the system:
\begin{equation}
\hat{\rho}(t)=\hat{\rho}_\mathrm{ss}+\sum_{j\geq1}\text{Tr}[\hat{l}_j^\dagger\hat{\rho}(0)]\hat{r}_j e^{\lambda_j t}.    
\end{equation}
Then, if there are $M$ eigenvalues whose real part (decay rate) is much smaller in absolute value than the rest, after an initial short transient, the dynamics is approximately contained in the manifold spanned by these eigenmodes  \cite{Macieszczak2016}:
\begin{equation}
\hat{\rho}(t)\approx    \hat{\rho}_\mathrm{ss}+\sum_{j=1}^M\text{Tr}[\hat{l}_j^\dagger\hat{\rho}(0)]\hat{r}_j e^{\lambda_j t}, \quad \text{for} \quad t\gg \frac{1}{|\text{Re}[\lambda_{M+1}]|}.
\end{equation}

The projection of a state on this metastable manifold is defined as:
\begin{equation}\label{projected_manifold}
\mathcal{P}\hat{\rho}=\hat{\rho}_\mathrm{ss}+\sum_{j=1}^M\text{Tr}[\hat{l}_j^\dagger\hat{\rho}(0)]\hat{r}_j,    
\end{equation}
and it is usually  insightful to compute expected values on this manifold  \cite{Rose2016,Cabot2021,Cabot2022}:
\begin{equation}\label{projected_average}
\langle \hat{O}\rangle_\mathcal{P}=\text{Tr}[\hat{O}\mathcal{P}\hat{\rho}].
\end{equation}
In fact, from Fig.~\ref{fig_SM_meta_2}(a) we see that $M=N/2$, which points out that the manifold defined by the projection $\mathcal{P}$ is related to the stationary states for each $S$ sector for the case $\gamma/\kappa=0$. Thus, the presence of weak local decay lifts the eigenvalue degeneracy at $0$ and leads to a metastable set of eigenmodes that are closely related to the old stationary states. This is confirmed when computing $\langle \hat{S}_\mathrm{y,z}\rangle_\mathcal{P}=\text{Tr}[\hat{S}_\mathrm{y,z}\mathcal{P}\hat{\rho}_{S_0}]$, as shown in color points in Fig.~\ref{fig_SM_meta_2}(b). The color indicates the value of $S_0$ of the initial state with zero magnetization $S_z$. We observe that $\langle \hat{S}_\mathrm{y,z}\rangle_\mathcal{P}$ are very close to the corresponding color lines, which are obtained from the stationary states in the case $\gamma/\kappa=0$. For comparison, in red-dashed line the true stationary state values are shown. In Fig.~\ref{fig_SM_meta_2}(d) we repeat the analysis but for $\gamma/\kappa=0.05$. We observe similar results for the considered $S_0$, despite the spectral gap only holding for a smaller set of eigenvalues. Moreover, these results are largely independent on the considered state within the particular $S_0$ sector, i.e. the initial magnetization $S_\mathrm{z}$ does not play an important role (not shown). In conclusion, the presence of these long-lived eigenmodes closely related to the stationary states for $\gamma/\kappa=0$ elucidates why the adiabatic ramps in Fig.~\ref{fig_SM_meta_1} follow closely the curves corresponding to $\gamma/\kappa=0$ instead of those belonging to $\gamma/\kappa>0$.

\begin{figure}[t!]
 \centering
 \includegraphics[width=1\columnwidth]{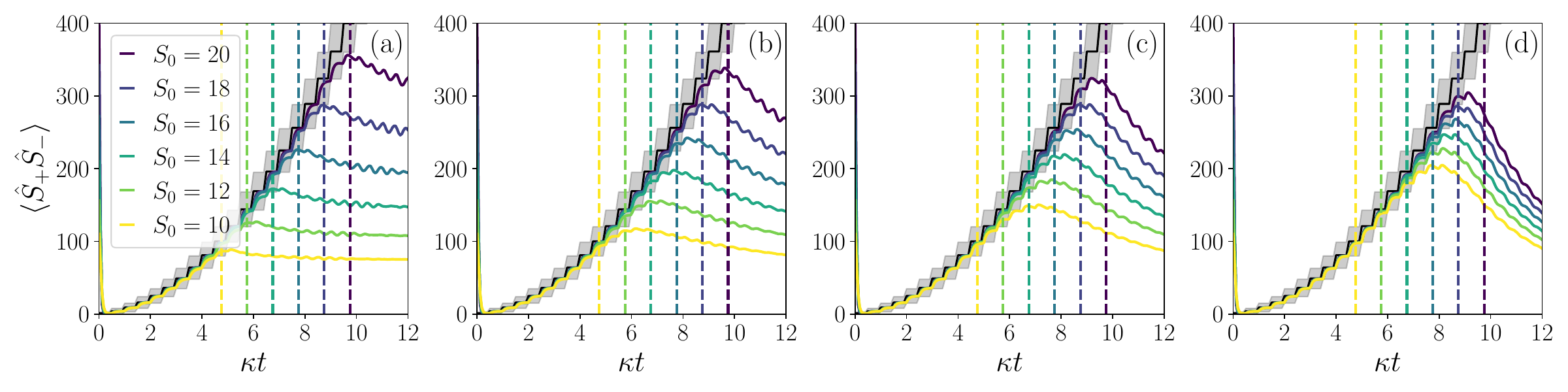}
 \caption{{\bf Adiabatic protocol in the presence of local losses for $N=40$.} (a) $\gamma/\kappa=0$, (b) $\gamma/\kappa=0.05$, (c) $\gamma/\kappa=0.1$, (d) $\gamma/\kappa=0.2$. Color lines correspond to the intensity at the average over infinite realizations, for initial conditions starting at different $S$ sectors (see legend). Parameters of the adiabatic ramp: $\kappa\Delta t=0.5$. The black solid line represents the universal curve for $\gamma/\kappa=0$, while the shadow corresponds to two standard deviations according to Eq. (\ref{moments_intensity}).}
 \label{fig_SM_adiabatic}
\end{figure}

Finally, in Fig.~\ref{fig_SM_adiabatic} we show the results of applying the adiabatic protocol  in the presence of local losses of various strengths, for several sectors $S$ and for $N=40$. In panel (a) we show the case of $\gamma/\kappa=0$; in panel (b) $\gamma/\kappa=0.05$; in panel (c) $\gamma/\kappa=0.1$; in panel (d) $\gamma/\kappa=0.2$. In all cases we keep $\kappa\Delta t=0.5$ and $\eta=1$, while the results are shown on the average over realizations (i.e. as computed from the master equation). We consider the system to be initialised in the state with total angular momentum $S_0\in\{10,12,14,16,18,20\}$ and fixed initial magnetization $S_z=0$.  We observe that the different curves still follow the universal curve (black line) until they depart from it at a $S_0$-dependent value. The value at which they deviate from the universal curve is not exactly the same as in the case $\gamma/\kappa=0$, the deviations being more significant for the cases belonging to a smaller initial $S_0$ and also for increasing $\gamma/\kappa$. Nevertheless, for the cases $\gamma/\kappa=0.05$ and $\gamma/\kappa=0.1$ [panels (b) and (d)], the curves belonging to different $S_0$ are still quite well resolved. This suggests that the proposed sensing protocol is robust against not too strong local decay channels, working better the larger $S_0$ is.

\section{Analysis of dissipative freezing}

\subsection{Master equation for multiple total angular momentum sectors}

In the study of dissipative freezing, we consider initial states containing superpositions and mixtures of different total angular momentum sectors. We can express them in the following form, generally containing coherences between different $S$ sectors:
\begin{equation}
\hat{\rho}(t)=\sum_{S,S'}\sum_{S_\mathrm{z},S'_\mathrm{z}}\rho^{S,S'}_{S_\mathrm{z},S'_\mathrm{z}}(t)|S,S_\mathrm{z}\rangle\langle S',S'_\mathrm{z}|.
\end{equation}
In terms of this parametrization, the master equation (\ref{ME_model1}) reads:
\small
\begin{equation}
\begin{split}
\partial_t \rho^{S,S'}_{S_\mathrm{z},S'_\mathrm{z}}=&-i\frac{\Omega}{2}\bigg[\sqrt{(S-S_\mathrm{z}+1)(S+S_\mathrm{z})}\rho^{S,S'}_{S_\mathrm{z}-1,S'_\mathrm{z}}(1-\delta_{S_\mathrm{z},-S})+\sqrt{(S+S_\mathrm{z}+1)(S-S_\mathrm{z})}\rho^{S,S'}_{S_\mathrm{z}+1,S'_\mathrm{z}}(1-\delta_{S_\mathrm{z},S})\\
&-\sqrt{(S'-S'_\mathrm{z}+1)(S'+S'_\mathrm{z})}\rho^{S,S'}_{S_\mathrm{z},S'_\mathrm{z}-1}(1-\delta_{S'_\mathrm{z},-S'})-\sqrt{(S'+S'_\mathrm{z}+1)(S'-S'_\mathrm{z})}\rho^{S,S'}_{S_\mathrm{z},S'_\mathrm{z}+1}(1-\delta_{S'_\mathrm{z},S'}) \bigg]\\
&+\kappa\bigg[\sqrt{(S+S_\mathrm{z}+1)(S-S_\mathrm{z})(S'+S'_\mathrm{z}+1)(S'-S'_\mathrm{z})}\,\rho^{S,S'}_{S_\mathrm{z}+1,S'_\mathrm{z}+1}(1-\delta_{S_\mathrm{z},S})(1-\delta_{S'_\mathrm{z},S'})\\
&-\frac{1}{2} \big((S+S_\mathrm{z})(S-S_\mathrm{z}+1)+(S'+S'_\mathrm{z})(S'-S'_\mathrm{z}+1)\big) \rho^{S,S'}_{S_\mathrm{z},S'_\mathrm{z}}\bigg].
\end{split}
\end{equation}
\normalsize
Vectorizing this equation, we observe that the Liouvillian is a block diagonal matrix and each possible pair of values $\{S,S'\}$ defines an independent block of it:
\begin{equation}
\mathcal{L}=\bigoplus_{S,S'}\mathcal{L}_{S,S'}.
\end{equation}
For finite $N$, only the blocks $\mathcal{L}_{S,S}$ contain a zero eigenvalue and thus a stationary state. In contrast, the coherences governed by $\mathcal{L}_{S,S'}$ with $S\neq S'$ decay in time, such that for long times $\rho^{S,S'\neq S}_{S_\mathrm{z},S'_\mathrm{z}}(t)=0$. In practice, if the initial state is  contained only in two different $S$ sectors ($S_{1,2}$), we only have to implement the corresponding blocks of the Liouvillian: $\mathcal{L}_{S_1,S_1}\oplus\mathcal{L}_{S_1,S_2}\oplus\mathcal{L}_{S_2,S_1}\oplus\mathcal{L}_{S_2,S_2}$. From this representation, we can also directly write down the stochastic master equation for the photocounting process.

\subsection{Partial dissipative freezing}

\begin{figure}[t!]
 \centering
 \includegraphics[width=0.8\columnwidth]{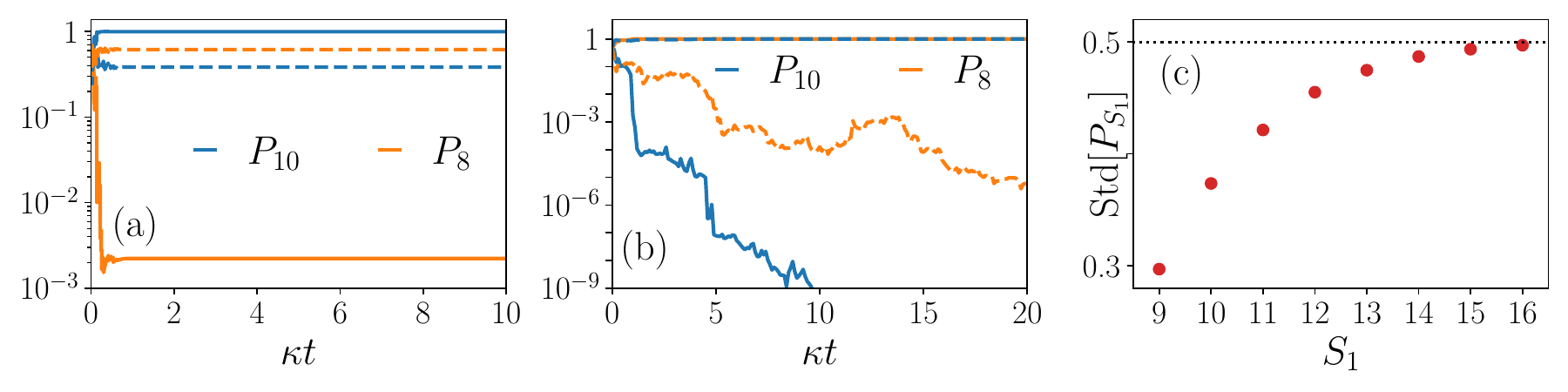}
 \caption{{\bf Dissipative freezing.} (a) Time evolution of the expected value of the projectors, $P_{S_{1,2}}$, for the initial state $|\Psi(0)\rangle$ with $S_1=10$ and $S_2=8$. Solid lines correspond to a realization in which the system mostly collapses into one sector, while dashed lines to a realization in which the weights on each sector remain similar. Here we fix $\omega/\kappa=S_2/2$ and $\eta=1$. (b) Time evolution of the expected value of the projectors, $P_{S_{1,2}}$, for $S_1=10$, $S_2=8$ and $\omega/\kappa=S_2$. Solid lines correspond to $\eta=1$ and  the initial state $|\Psi(0)\rangle$. Dashed lines correspond to $\eta=0.5$ and the initial condition $\hat{\rho}(0)$. (c) Standard deviation over stochastic realizations of $P_{S_1}$ at $\kappa t=10$, varying $S_1$ and keeping $S_2=8$ for the initial state $|\Psi(0)\rangle$. The average has been performed over $5\cdot10^4$ realizations. Here we fix $\omega/\kappa=S_2/2$ and $\eta=1$.}
 \label{fig_SM_freezing_1}
\end{figure}

In the overdamped regime dissipative freezing occurs only partially as the system is able to reach a stationary state that still spans more than one symmetry sector. The weights on the different sectors vary during the transient to this stationary state, and  their final value is random for each realization [see Fig. \ref{fig_SM_freezing_1} (a)]. For this reason we refer to this case as partial freezing.

{\it Stationary state of the photocounting process. --} In order to understand why partial freezing occurs, we need to recall some properties of the stationary state in this regime. Within each sector, the stationary state is pure and it satisfies to a very good approximation the eigenvalue like equation \cite{Cabot2023}:
\begin{equation}
\hat{S}_-|\Psi_{S}\rangle\approx -i\frac{\omega}{\kappa}|\Psi_{S}\rangle.  
\end{equation}
The eigenvalue $\omega/\kappa$ is independent of $S$, as long as the corresponding sector is well in the overdamped regime. Notice that this is no longer the case in the crossover region, when $\omega$ approaches $ \omega_c(S)$ from below. However, the relative size of the crossover region diminishes with $S$, as shown in Ref. \cite{Cabot2023}. The action of the effective Hamiltonian on this state gives:
\begin{equation}
\hat{H}_\mathrm{eff}|\Psi_{S}\rangle\approx -i\frac{\omega^2}{2\kappa}|\Psi_{S}\rangle.
\end{equation}
This independence on $S$ of the action of $\hat{H}_\mathrm{eff}$ and $\hat{S}_-$ on $|\Psi_{S}\rangle$ prevents the general occurrence of dissipative freezing. This can be understood by considering the following superposition state between two total angular momentum sectors $S_1$ and $S_2$:
\begin{equation}
|\Psi\rangle=C_{S_1}|\Psi_{S_1}\rangle+C_{S_2}|\Psi_{S_2}\rangle.    
\end{equation}
For simplicity we assume ideal photocounting $\eta=1$, although we recall that the same arguments hold for $\eta<1$. Assuming the first jump occurs at $t_1$, the (unnormalized) state before the first count is:
\begin{equation}
e^{-i\hat{H}_\mathrm{eff}t}|\Psi\rangle=\big(C_{S_1}|\Psi_{S_1}\rangle+C_{S_2}|\Psi_{S_2}\rangle\big)e^{-\omega t/(2\kappa)}.    
\end{equation}
At $t_1$, we apply the jump operator and renormalize the state obtaining:
\begin{equation}
|\Psi(t_1)\rangle= C_{S_1}|\Psi_{S_1}\rangle+C_{S_2}|\Psi_{S_2}\rangle,   
\end{equation}
which shows that this family of states is invariant under the photocounting evolution. This argument can be generalized to any number of counts and to any superposition of $|\Psi_S\rangle$ of different sectors, as long as all the sectors are well into the overdamped regime. 

{\it Partial freezing. --} The photocounting process eventually reaches a state in which each of the initially populated sectors have reached their corresponding $|\Psi_S\rangle$, since this state is an attractor of the dynamics for each sector \cite{Cabot2023}. Once there, the state is invariant to time evolution. This is illustrated in Fig.~\ref{fig_SM_freezing_1}(a), in which we show two realizations of the dynamics with initial state $|\Psi(0)\rangle=\frac{1}{\sqrt{2}}(|S_1,0\rangle+|S_2,0\rangle)$. In both cases, after a short transient, the expected values of the projectors are constant in time, indicating partial freezing. This is in stark contrast to the standard dissipative freezing \cite{Tindall2023} [e.g. Fig.~\ref{fig_SM_freezing_1}(b)]. Nevertheless, we notice that the weights between the different sectors can be very different and the system might reach a final state in which it has mostly collapsed into one sector [e.g. Fig.~\ref{fig_SM_freezing_1}(a) solid lines].

In fact, we observe that the final state resides mostly in one sector when $S_1$ and $S_2$ are not close to each other. In order to analyze this in more detail, we compute the standard deviation over stochastic realizations of $P_{S_{1,2}}$ after a long time and varying $S_1-S_2$ [see Fig.~\ref{fig_SM_freezing_1}(c)]. When considering the initial state $|\Psi(0)\rangle$, this standard deviation tends to $0.5$ when the final state resides in just one sector, while it is smaller otherwise. From Fig.~\ref{fig_SM_freezing_1}(c) we observe that this is the case as $S_1-S_2$ increases. Thus for large differences in the symmetry sectors, the final state mostly resides into one symmetry sector and for practical means there is no difference with the case of standard dissipative freezing.

Finally, in Fig.~\ref{fig_SM_freezing_1}(b), we consider the case in which the sector with smallest $S$ is at the crossover, i.e. $\omega=\kappa S_2$. Both for an initial superposition $|\Psi(0)\rangle$ with $\eta=1$, or for an initial mixture $\hat{\rho}(0)=\frac{1}{2}(|S_1,0\rangle\langle S_1,0|+|S_2,0\rangle\langle S_2,0|)$ with $\eta=0.5$, standard dissipative freezing occurs and the state collapses randomly into one of the $S$ sectors. Thus, when applying the adiabatic ramp of the Rabi frequency, the system eventually collapses (randomly) into one of the sectors, as it eventually reaches a crossover region in which full collapse occurs.

\end{document}